\documentclass[reprint]{revtex4-2}

\usepackage[utf8]{inputenc}
\usepackage{amsmath}
\usepackage{amsfonts}
\usepackage{amssymb}
\usepackage{latexsym}
\usepackage{mathrsfs}
\usepackage{mathtools}
\usepackage{braket}		%Bra Ket Notation
\usepackage{graphicx}
\usepackage{color}
\usepackage{xcolor}
\usepackage{slashed}
\usepackage{twistor}
\usepackage[all]{xy}
\usepackage{tikz-cd}
\usepackage{hyperref}
\hypersetup{
    colorlinks = true,
    urlcolor = {brown},
}

\newcommand{\veps}{\varepsilon}

\newcommand{\h}{\mathbf{h}}

\renewcommand{\d}{\mathrm{d}}

\newcommand{\br}[1]{\overline{#1}}

\newcommand{\g}{\mathfrak{g}}

\renewcommand{\dal}{{\dot\alpha}}

\newcommand{\AdS}{\text{AdS}}
\newcommand{\ST}{\mathbb{ST}}

\begin{document}

\title{Chiral higher-spin theories from twistor space}

\author{Lionel Mason,}
\affiliation{The Mathematical Institute, University of Oxford, United Kingdom}
\email{lmason@maths.ox.ac.uk}

\author{Atul Sharma}
\affiliation{Center for the Fundamental Laws of Nature, Harvard University, Cambridge, USA}
\email{atulsharma@fas.harvard.edu}

\begin{abstract}
    We reformulate chiral higher-spin Yang-Mills and gravity on $\R^4$ as `CR-holomorphic' theories of Chern-Simons type;  in the most general case, these are  Moyal deformed to become non-commutative.  They are defined on the space of non-projective twistors of unit length. These spaces carry  $S^7$, $S^3\times \R^4$  or AdS$_{3+4}$ metrics but are also endowed with a \emph{Cauchy-Riemann structure}, an odd-dimensional analogue of a complex structure, with respect to which the theories are holomorphic. They are circle bundles over standard projective twistor spaces and the higher  spin fields arise naturally as Kaluza-Klein modes.  We give a perturbative analysis to identify the spectrum and three-point vertices on spacetime and, for flat space, in momentum space. These vertices can have helicities  $(+++)$, $(++-)$ or $(+--)$, but are nevertheless all of $\overline{\text{MHV}}$ type in the sense that they are supported on momenta with proportional anti-self-dual spinors. On reduction to spacetime, there are higher valence vertices but these appear to be a gauge artifact. Further  generalizations are discussed. 
\end{abstract}

\maketitle

\section{Introduction}

Massless higher-spin theories arise naturally as certain limits of string theory, and have become increasingly important in holography where they emerge naturally as the bulk duals of boundary conformal field theories with conserved higher-spin currents. In four dimensions, they contain chiral sectors that are both building blocks of parity invariant theories \cite{Adamo:2022lah} and are thought to possess integrability properties that make them, and perhaps their holographic duals, more accessible  \cite{Sharapov:2022awp,Sharapov:2022wpz, Aharony:2024nqs}.
Chiral higher-spin theories were first constructed in \cite{Ponomarev:2016lrm} based on the analysis in \cite{Metsaev:1991nb, Metsaev:1991mt}.  These have further truncations to integrable self-dual Yang-Mills and gravity theories  \cite{Ponomarev:2017nrr} with spacetime actions found in \cite{Krasnov:2021nsq}; see the table in \cite{Monteiro:2022xwq}. 

In twistor theory, the Penrose transform encodes fields of arbitrary spins in terms of cohomology classes of different homogeneities on projective twistor space $\PT$ \cite{Eastwood:1981jy}. This has led a number of authors over the years to consider twistor formulations of higher-spin theories such as  \cite{Adamo:2016ple} for  conformal higher-spin theories, and \cite{Herfray:2022prf,Tran:2021ukl} for some of the theories described above.  In these formulations, the higher-spins naturally knit together to give theories on non-projecive twistor space $\T=\C^4$ that no longer descend to the projective space.  However, the theories of \cite{Herfray:2022prf,Tran:2021ukl} do not give all the theories listed in \cite{Monteiro:2022xwq} nor incorporate all the relevant interactions.  Here we give a natural twistor formulation that gives maximal chiral theories and a fully nonlinear approach to the subject.

In this paper, we give a simple twistor action of holomorphic (or more accurately CR) Chern-Simons type for  chiral higher-spin theories of all helicities.   It lives on the space of twistors of unit length adapted to spacetimes of Euclidean signature.  This has three versions depending on whether we use $\SU(4)$, $\SU(2,2)$, or degenerate rank-two norms, corresponding to choices of positive, negative or vanishing cosmological constant. The first two correspond to theories on $S^7$ or on $\AdS_7$ of  signature $(3,4)$. The degenerate  flat case is a $\C^2$ fibre bundle over $S^3$.  These are endowed with \emph{Cauchy-Riemann} (CR) structures, the structure that a real hypersurface inherits from its embedding in a complex manifold, here $\C^4$.  These admit holomorphic 4-forms which allow us to introduce CR analogues of Chern-Simons theories on these spaces. 

We first discuss chiral higher-spin theories of Yang-Mills and gravity. In section \S\ref{Moyal-defs}, we also give the non-commutative Moyal deformation that has limits onto both the previous theories.  
We give a perturbative analysis of each of these theories in turn,  reducing to perturbative theories on spacetimes.  We will see that the spectrum includes all spins, both integer and half-integer, although various truncations are possible. We characterize their 3-point vertices. They  are all of $\overline{\text{MHV}}$ type in the sense that they are supported on momenta whose undotted spinors are all  proportional. They  must therefore be represented in terms of self-dual spinor inner products, i.e., square bracket in spinor-helicity notation. One can write three-point $\overline{\text{MHV}}$ vertices with any helicity assignment, but we only obtain those whose sum is 1 for the chiral higher-spin Yang-Mills, 2 for chiral higher-spin gravity, and extending to all positive integers after Moyal deformation. On spacetime there are vertices of all orders, but these are artifacts of the spacetime gauge. We briefly discuss a number of possible further developments in the discussion section.

%%%%%%%%%%%%%%%%%%%%%%%%%%%%%%%%%
%%%%%%%%%%%%%%%%%%%%%%%%%%%%%%%%%

\section{Twistor geometry}

A non-projective twistor $Z^A=(\lambda_{\alpha},\mu^{\dot\alpha})$ is an element of $\T=\C^4$, where $\al=0,1$, $\dal=\dot0,\dot1$ are $\SL_2(\C)$ spinor indices. 
We will work with Euclidean signature spacetimes.  This allows for a quaternionic complex conjugation 
\begin{equation}
    \hat\lambda_{\alpha}=(-\br{\lambda_1},\br{\lambda_0})\, , \qquad \hat{\hat{\lambda}}_\alpha=-\lambda_\alpha\,,
\end{equation}
with a similar definition for $\hat\mu^{\dot\alpha}$ inducing the conformally invariant quaternionic conjugation $Z^A\mapsto \hat Z^A$ on $\T$.  We will use the spinor-helicity notation
\begin{equation}
\langle ab\rangle\vcentcolon=    \varepsilon^{\alpha\beta}a_{\alpha}b_{\beta}\, , \qquad [ab]\vcentcolon=\varepsilon_{\dot\alpha\dot\beta}a^{\dot\alpha}b^{\dot\beta}
\end{equation}
for contractions constructed from $\SL_2(\C)$-invariant Levi-Civita symbols $\veps^{\al\beta}$, $\veps^{\dal\dot\beta}$.

We introduce norms on $\T$ via choices of constant curvature conformal scale on spacetime. These are encoded in a skew ``infinity twistor'' $I_{AB}=I_{[AB]}$, 
\begin{equation}
    I_{AB}\,\d Z^A\wedge\d Z^B=\langle \d\lambda\wedge \d\lambda\rangle + \Lambda [\d\mu\wedge \d\mu]\, ,
\end{equation}
with $\Lambda$ the cosmological constant.  This gives the norm
\begin{equation}
    |Z|^2=I_{AB}Z^A\hat Z^B = |\lambda|^2+\Lambda|\mu|^2
\end{equation}
defining an $\SU(4)$ inner product for $\Lambda>0$, $\SU(2,2)$ for $\Lambda<0$, and degenerate of rank two for $\Lambda=0$.

Our master geometry will be the 7-manifold of unit norm twistors
\be
\ST = \{Z^A\in\T\,:\,|Z|=1\}\, .
\ee
Projective twistor space $\PT\subset\CP^3$ is obtained as the quotient of $\ST$ by the $\U(1)$ action $Z^A\sim\e^{i\theta}Z^A$, $\theta\in S^1$. $\ST$ is $S^7$ when $\Lambda$ is positive and $\AdS_7$ when  negative. In all cases we have the Atiyah, Hitchin, Singer (AHS) fibration \cite{Atiyah:1978wi} which non-projectively fibres $\T-\{0\}\rightarrow \M$ with fibre $\C^2-\{0\}$, where $\M$ is  
spacetime of Euclidean signature, which we take to be a region in $S^4$, $\R^4$ or hyperbolic 4-space $\mathbb{H}^4$ depending on our choice of $\Lambda$.  Locally on $\M$  we use coordinates $x^{\alpha\dot\alpha}$ and metric 
\begin{equation}
\d s^2= \Omega^2 \d x^{\alpha\dot\alpha} \d x_{\alpha\dot\alpha}\, , \qquad \Omega=(1+\Lambda x^2)^{-1}\, .
\end{equation}
We use $(x^{\alpha\dot\alpha},\sigma_\alpha)$ as coordinates on the anti-self-dual spin-bundle $\mathbb{S}^-\rightarrow \M$ which is identified with the non-projective AHS fibration to $\M$ by setting  
\begin{equation}
(\lambda_\alpha,\mu^{\dot\alpha})=\sqrt{\Omega}\,(\sigma_\alpha,x^{\alpha\dot\alpha}\sigma_\alpha)\, , \quad x^{\alpha\dot\alpha}= \frac{\mu^{\dot\alpha}\hat\lambda^\alpha- \hat\mu^{\dot\alpha}\lambda^\alpha}{\langle\lambda\hat\lambda\rangle}
\end{equation}
In these coordinates $\ST=\{(x,\sigma)\,|\, \langle\sigma \hat\sigma\rangle=1\}$.

The Penrose transform encodes spacetime fields in terms of complex analytic cohomology, but here, 
as a real codimension-one submanifold of the complex manifold $\T$, $\ST$ only inherits a Cauchy-Riemann structure.  However, this too can be used to define cohomology.  In particular, its complexified tangent bundle admits a 3-dimensional  sub-bundle $T^{0,1}$ defined to be the sub-bundle of $(0,1)$-vectors on $\T$ that are tangent to $\ST$.  The bundle of $(p,0)$-forms on $\T$ restricts to give the ${4\choose p}$-dimensional sub-bundle $\Omega^{p,0}$ of the complex p-forms $\Omega^p$ on $\ST$ that annihilates $T^{0,1}$.  

The boundary  $(p,q)$-forms $\Omega^{p,q}_b$ on $\ST$ are 
defined as quotients with the first example    
\begin{equation}
\Omega^{0,1}_b=\Omega^1/\Omega^{1,0}=(T^{0,1})^*   \, ,
\end{equation}
giving a 3-dimensional bundle. 
We similarly define  
\begin{equation}
\Omega^{p,q}_b = \Omega^{p+q}/  \{\Omega^{p+1,0}\wedge\Omega^{q-1}\}\, .
\end{equation}  
The integrability of $T^{0,1}$ means dually that $\d(\Omega^{1,0})\subset \Omega^{1,0}\wedge \Omega^1$ so that it is consistent to define the tangential Cauchy-Riemann operator $\dbar_b:\Omega_b^{p,q}\rightarrow \Omega_b^{p,q+1}$, where  $\d^2=0$ implies also that $\dbar_b^2=0$.  

Adapting our forms to the AHS $(x,\sigma)$-coordinates  we find on $\ST$   the basis
for $(0,1)$-forms  
\begin{equation}
(\bar e^0,\bar e ^{\dot\alpha}):=(\langle\hat\sigma D\hat\sigma\rangle\, ,  \hat \sigma_\alpha e^{\alpha\dot\alpha})\, , \label{form-basis}
\end{equation}
where $e^{\alpha\dot\alpha}=\Omega\, \d x^{\alpha\dot\alpha}$,  and $D\hat \sigma_\alpha$ incorporates the undotted spin connection.  Dually we have the  vector fields spanning $T^{0,1}$
\begin{equation}
(\bar\eth,\bar\nabla_{\dot\alpha}):=\left(\left\langle\sigma \frac{\p}{\p\hat\sigma}\right\rangle, \sigma^\alpha\nabla_{\alpha\dot\alpha}\right) 
\end{equation}
 where $\nabla_{\alpha\dot\alpha}$ is the horizontal lift of the vector fields dual to $e^{\alpha\dot\alpha}$ to the spin bundle. We normalize $\langle\sigma\hat \sigma \rangle=1$ so that $\hat\sigma$ and hence $\bar e^{\dot\alpha}$ has weight $-1$ in $\sigma_\alpha$ and $\bar e^0$ weight $-2$ and with these weights (or line bundle assignements), these forms and vector fields descend to the projective space.

%%%%%%%%%%%%%%%%%%%%%%%%%%%%%%%%%
%%%%%%%%%%%%%%%%%%%%%%%%%%%%%%%%%

\section{Chiral higher-spin Yang-Mills}

Our first theory is a twistor action for the theory labeled gl-chs$(0)$ in table 1 of \cite{Monteiro:2022xwq}, where `gl' stands for gluonic and `chs' stands for chiral higher-spin. It is given by a 7D CR-Chern-Simons action
\be\label{7dCS}
S[a] = \frac{1}{2\pi}\int_{\ST}\d^4Z\wedge\tr\bigg(a\wedge \d a + \frac23\,a^3\bigg)\,,
\ee
where $a\in\Omega^{0,1}_b(\ST,\g)$ is a partial connection valued in the Lie algebra $\g$ of a gauge group $G$.  The equations of motion will then  be 
\begin{equation}
F_a:=(\dbar_b +a)^2 = \dbar_b a+ \half\, [a,a]=0\label{YM-eom}
\end{equation}
so that $\dbar_b+a$ is a CR connection.  
%We can similarly consider $B$-$F$ actions with  $B\in\Omega^{0,1}_b(\ST,\g)$ 
%\be\label{7BF}
%S[a] = \frac{1}{2\pi}\int_{\ST}\d^4Z\wedge\tr\big(B\wedge F_a\big)\,,
%\ee
%with the same equations of motion for $a$ and $\dbar_bB+[a,B]=0$.
% These will descend to theories of chiral higher-spin Yang-Mills on spacetime.

Our first step in obtaining the perturbative spacetime action is to decompose the fields into Fourier modes on  the $S^1$ fibres of $S^1\hookrightarrow \ST\rightarrow \PT$; this will yield an action on projective twistor space.  The $S^1$ fibres are generated by the vector field $\p_\theta:=i(\Upsilon-\br\Upsilon)$, where 
\begin{equation}
    \Upsilon=Z^A\frac{\p}{\p Z^A}=\left\langle\sigma \frac{\p}{\p\sigma}\right\rangle\, .
\end{equation}
We can decompose $a$ into Fourier modes in the fibre $S^1$'s 
\begin{equation}
    a=\sum_{n\in \Z} a_n\, , \quad  \mathcal{L}_{\p_\theta %\Upsilon-\br\Upsilon
    } a_n=n a_n\, .
\end{equation}
This decomposition is compatible with the $\dbar_b$-operator, because it is essentially a Laurent expansion based on the unit circle inside the $\C^*$ fibres of $\T-\{0\}\rightarrow \CP^3$.  This   is the holomorphic line bundle $\cO(-1)$  and so can be trivialized locally holomorphically. So we find 
\begin{equation}
\dbar_b a= \sum_{n\in \Z} \dbar a_n
 \end{equation}
where now on the right hand side the $a_n\in \Omega^{0,1}(n)$, where we have abbreviated $\Omega^{0,1}(n)\vcentcolon=\Omega^{0,1}\otimes \cO(n)$ and $\dbar$ is the standard $\dbar$-operator on sections of that line bundle.

Plugging these coordinates and Fourier expansions into \eqref{7dCS} and performing the integral over $S^1$ yields the twistor action
\begin{multline}\label{HSYM-incpts}
    S[a] =  \int_{\PT}\D^3Z\wedge\tr\bigg(\sum_n a_{-n-4}\wedge\dbar a_n\\
     + \frac23\sum_{\ell+m+n=-4}a_\ell\wedge a_m\wedge a_n\bigg)\,,
\end{multline}
where $\D^3Z\in \Omega^{3,0}(4)$ is the projective volume form 
\be
\D^3Z = \frac{1}{3!}\,\veps_{ABCD}\,Z^A\,\d Z^B\wedge\d Z^C\wedge\d Z ^D\,,
\ee
with $\veps_{ABCD}$ being the 4-dimensional Levi Civita symbol.

Our original Chern-Simons action has full gauge freedom on $\ST$. Decomposing this into modes we obtain infinitesimally
\begin{equation}
\delta a_n=\dbar \chi_n +\sum_{l\in\Z} [a_{l},\chi_{n-l}]\, .
\end{equation}
These are compatible with the equations of motion
\begin{equation}
\dbar a_n + \sum_k [a_k, a_{n-k}]=0\, . \label{gauge1}
\end{equation}

Thus, in the linear approximation, we have $\dbar$-closed $(0,1)$-forms $a_n$ modulo $\dbar$-exact $(0,1)$-forms of each homogeneity defining cohomology classes $[a_n]\in H^{0,1}(\PT,\CO(n))$.  It is well-known that via the Penrose transform \cite{Eastwood:1981jy}, these correspond to massless fields of helicity $h_n=(n+2)/2$ and so we have an interacting theory of Lie-algebra-valued fields of all helicities.  The cubic vertices therefore encode interactions between three particles whose helicities satisfy $h_1+h_2+h_3=1$.

\subsection{Reduction to spacetime action}

To reduce our twistor space action  to  spacetime,  we follow the strategy of \cite{Mason:2005zm,Boels:2006ir} and start by gauge fixing up the fibres of the AHS fibration $p:\ST\rightarrow \M$.
%and integrate out fields that appear linearly in the action.
 This follows \cite{Woodhouse:1985id} and is sufficient for a perturbative analysis.  Using the basis \eqref{form-basis}, we  write
\begin{equation}
a_n= a_{n0}\bar e^0
%\langle \hat \sigma d\hat \sigma\rangle
 + a_{n\dot\alpha}\bar e^{\dot\alpha}\,  ,
\end{equation}
where now $a_{n0}$ has weight $n+2$, and $a_{n\dot\alpha}$ weight $n+1$.
 For  $n\geq -1$ there is no cohomology on $\CP^1$.
 Thus, if we supplement the action with  a harmonic gauge condition on the $\CP^1$-fibres, the fibre component $a_{n0}$ of $ a_n$  will vanish  for $n\geq -1$. 
 %\footnote{ This is sufficient for the perturbative case; more generally, an elementary inductive argument decreasing from large $n$  suffices if the $a_n$ are assumed to vanish for sufficiently large $n$ and are assumed sufficiently small.}
 %In harmonic gauge up the AHS fibres, we have $a_{n0}=0$ for $n\geq -1$ and 
 On the other hand, for $n\leq-2$ we have  nontrivial cohomology up the AHS fibres which in harmonic gauge is fixed to be \cite{Woodhouse:1985id} \begin{equation}
a_{-n-2\, 0}=\phi_{\alpha_1\ldots\alpha_n}\hat\sigma^{\alpha_1}\ldots \hat\sigma^{\alpha_n}  \, .\label{neg-harm}
\end{equation}
Here $\phi(x)_{\alpha_1\ldots\alpha_n}$ depends only on $x$, reflecting the fact that $H^{0,1}(\CP^1,\cO(-n-2))=\C^{n+1}$, realized as spinors with $n$ symmetric indices.

We now attempt to integrate the  equations of motion up the AHS fibres.  
%observe that for $n  \geq -1 $, $a_{-n-1  \dot\alpha}$ can only appear linearly in our action as only two of the factors of the cubic terms can be negative, and one of those must provide the  $\langle \hat \sigma d\hat \sigma\rangle$ factor. We can therefore regard it as a Lagrange multiplier, integrate it out imposing the equation of motion
In harmonic gauge we have
\begin{equation}
\bar\eth a_{n\dot\alpha}+\sum_{k\geq 2}[a_{-k\, 0}, a_{n +k\,\dot \alpha}]= \begin{cases}
    0\, , \;&n\geq -1\, ,\\\bar \nabla_{\dot\alpha} a_{n\, 0}\, , &n\leq -2\,.
\end{cases}\label{hol-a-dot-eq}    \end{equation}
%\bar\eth a_{n\,\dot\alpha} +\sum_{k\geq 2}[a_{-k\, 0},a_{n+k\, \dot\alpha}] &= \label{hol-a-dot-minus}
We solve these iteratively. For $n\geq -1$, $\bar \eth $ has a kernel, so the solutions naturally decompose as
\begin{equation}
a_{n\dot\alpha}=a_{n\dot\alpha}^h+
    a_{n\dot\alpha}^i\, , \qquad \bar \eth a^h_{n\dot\alpha}=0\, . 
\end{equation}
where the homogeneous term is holomorphic 
\begin{equation}
a_{n\dot\alpha}^h=\rho(x)^{\alpha_1\ldots \alpha_{n+1}}_{n \,\dot\alpha} \sigma_{\alpha_1}\ldots \sigma_{\alpha_{n+1}}\, ,\label{pos-harm}
\end{equation}
with $\rho(x)_{\dot\alpha}^{\alpha_1\ldots \alpha_{n+1}}$ now depending only on $x$. The residual gauge freedom is holomorphic in $\sigma_\alpha$, leading to 
\begin{equation}
    \delta\rho_{\dot\alpha}^{\alpha_2\ldots \alpha_n}= \nabla^{(\alpha_2}_{\dot\alpha}\gamma^{\alpha_3\ldots \alpha_n)}\, . \label{gauge}
\end{equation}
 The inhomogeneous term is defined by 
\begin{equation}
    a^i_{n\dot\alpha}=-\bar\eth^{-1} \left(\sum_{k\geq 2}[a_{-k\, 0}, a_{n +k\,\dot \alpha}] +\bar \nabla_{\dot\alpha} a_{n\, 0}\right)\, . \label{inhom}
\end{equation}
The second term on the RHS vanishes for $n\geq -1$ so that in this range $a^i_{n\dot\alpha}$ is already quadratic.
These can be solved via spherical harmonics up the AHS fibres following \S4.15 of \cite{Penrose:1984uia}. 
We choose the inverse $\bar\eth^{-1}$ so that $a^i_{n\dot\alpha}$ integrates to zero over the AHS fibres  against  expressions of the form \eqref{neg-harm} but  of weight $-n-3$.

We can now evaluate  the action  in the quadratic approximation as the sum over $n\geq 0$ of
\begin{align}
    &S[a]^{2}_n= \int_{\PT}  \D^3Z  \wedge \tr (a_{-n-2}\wedge \dbar a_{n-2})_2\, ,\\
    &= \int_{\PT}  \d^6Z  \wedge \tr (a_{-n-2\, 0}\bar\nabla^{\dot\alpha}a_{n-2\dot\alpha} + a_{-n-2}^{\dot\alpha}\bar\eth a_{n-2\dot\alpha}) \label{quad-action}
\end{align}
Here $\d^6Z=i\,\D^3Z\wedge \D^3\bar Z/|\sigma|^8$. 
%For $n\geq 1$ our choice of $\bar\eth^{-1}$ to define  $a^i_{n\, \dot\alpha}$  ensures that it projects to zero in the quadratic part of the action, so that it  requires just $a_{n-2\, \dot\alpha}^h$. As  there is no $\bar e^0$-term in $\dbar a_{n-2}$, the only contribution from $a_{-n-2}$ will be $a_{-n-2\, 0}\bar e^0 $ so 
To quadratic order, we can replace $a_{n-2}^{\dot\alpha}$ by its homogeneous part which for $n>0$ is holmorphic up the fibres, so that the second term vanishes.  
Now, integrating out the AHS fibres we obtain the spacetime kinetic term
\begin{equation}
    S[a]^{2}_n= \int \Omega^4\,\d^4x\;  \tr(\phi_{\alpha_1\ldots \alpha_n} \nabla^{\alpha_1\dot\alpha} \rho_{\dot\alpha}^{\alpha_2\ldots \alpha_n})  \, .
\end{equation}
This is a standard action for helicity $\pm n/2$ fields with field equations
\begin{equation}
\nabla^{\alpha_a\dot\alpha}\phi_{\alpha_1\ldots \alpha_n}=0=\nabla^{\dot\alpha (\alpha_1} \gamma_{\dot\alpha}^{\alpha_2\ldots \alpha_n)}    \, .
\end{equation}
Here $\rho_{\dot\alpha}^{\alpha_2\ldots \alpha_n}$ is a potential modulo  gauge \eqref{gauge}.
 The corresponding 
field is defined by
\begin{equation}
    \tilde \phi_{\dot\alpha_1\ldots\dot\alpha_n}=\nabla_{\alpha_2(\dot\alpha_2|}\nabla_{\al_3|\dal_3|}\ldots \nabla_{\alpha_n|\dot\alpha_{n}}\rho_{\dot\alpha_1)}^{\alpha_2\al_3\ldots \alpha_n}\, .
\end{equation} 

In the $n=0$ case, $a_{-2\dot\alpha}$ has no homogeneous term, and  in the linear approximation \eqref{hol-a-dot-eq} is solved by
\begin{equation}
    a_{-2\, \dot\alpha}= \nabla_{\dot\alpha} \phi(x)\, , \qquad \nabla_{\dot\alpha}:=\hat\sigma^\alpha \nabla_{\alpha\dot\alpha}
    %-\bar \nabla_{\dot\alpha} \bar\eth a_{n-2\, 0}
    %\phi_{\alpha_1 \ldots \alpha_n}\sigma^{\alpha_1}\hat\sigma^{\alpha_2}\ldots\hat \sigma^{\alpha_n} 
    \, .
\end{equation}  
 Integrating out the fibre, we land on the standard scalar massless field action. 
Thus we land on  spacetime actions for our spectrum of massless fields for each helicity.

\subsection{Three-point vertices}
The 3-point vertices in \eqref{HSYM-incpts} come in two types, with, two positive helicities and one negative helicity or vice versa.   In the first case, the interaction term is easily evaluated as 
\begin{equation}
    V^{++-}=\int \d^4\mu_\M \, \tr \left( \phi_{\alpha_1\ldots \alpha_n} \rho^{\alpha_1\ldots \alpha_l}_{l\,\dot\alpha}\rho_m^{ \alpha_{l+1}\ldots \alpha_n \dot\alpha}\right)\, .\label{spt++-}
\end{equation}
We also have vertices $ V^{--+}$ given by
\begin{equation}
   V^{--+}=\int \d^4\mu_\M \,\tr(\rho_{l\, \dot\alpha}^{\alpha_1\ldots \alpha_l}\phi_{\alpha_1\ldots\alpha_m}\nabla_{\alpha_{m+1}}^{\dot\alpha} \phi_{\alpha_{m+2}\ldots\alpha_l}) \, . \label{spt--+}
\end{equation}
Equivalent terms arise also from inserting $a^i_{n\dot\alpha}$ into \eqref{quad-action}.

When $\Lambda=0$ these vertices give simple formulae  on momentum space where it is easily seen that they are of $\overline{\text{MHV}}$ type. These are obtained by inserting the Penrose transform of momentum eigenstates into the vertices and integrating.   For momentum $k^{\alpha\dot\alpha}=\kappa^\alpha\tilde \kappa^{\dot\alpha}$ we have \begin{equation}
    a_n= \int_{\C^*} \frac{\d s}{s^{n+1}} \, \bar\delta ^2( \kappa -s\lambda) \, \e^{is[\tilde \kappa \mu]} \, .\label{mom-estate}
\end{equation}
Taking complex momenta, and the integrations over a nearly Lorentzian slice in the spacetime integral, the delta functions force the undotted momentum spinors to  be proportional giving our characterization of $\overline{\text{MHV}}$.\footnote{MHV will have  $h_1+h_2+h_3=-1$, proportional dotted momentum spinors,  and corresponding angle-bracket expressions.}  For helicities $h_i$, $i=1,2,3$ we have $h_1+h_2+h_3=1$ and
\begin{equation}
V(h_1,h_2,h_3)  =[1\,2]^{a_3}[2\,3]^{a_1}[3\,1]^{a_2} \delta^4(\sum_i   k_i)\label{3pt-mom}
\end{equation}
with $a_1=2h_2+2h_3-1$ and cyclic. 
%a+b=2h_2, b+c=2h_3, c+a=2h_1. 2h_1+2h_2=a+1

There are superficially many more spacetime vertices because the inhomogeneous terms  \eqref{inhom} must be substituted  into the cubic Chern-Simons vertices and kinematic terms leading to further hierarchies of vertices. This gives further cubic terms together with quartic and higher vertices.  These do not appear on twistor space nor on momentum space, which suggests that such vertices of  valence higher than three are spacetime gauge artifacts.  See \cite{Ren:2022sws} for related phenomena. 

%However, this one is harder to solve with the 2nd term which appears for the first time from higher-spin considerations and is quite non-trivial in general. 

%%%%%%%%%%%%%%%%%%%%%%%%%%%%%%%%%
%%%%%%%%%%%%%%%%%%%%%%%%%%%%%%%%%

\section{Chiral higher-spin gravity}

%Let $\CO(n)\to\ST$ be the $\R_{>0}$ bundle whose sections are functions of homogeneity $n$ under positive rescalings of $Z^A$. 
Our next theory is the one labeled chs(0) in table 1 of \cite{Monteiro:2022xwq}. Following \cite{Mason:2007ct}, we take our chiral higher-spin gravitational twistor action to be the Poisson Chern-Simons theory
\be
S[\h] = \int_{\ST}\d^4Z\wedge\bigg(\h\wedge\dbar_b \h + \frac13\,\h\wedge\{\h,\h\}\bigg)\,,
\ee
where $\h\in\Omega^{0,1}(\ST)$ %$,\CO(2))$, 
 and we have introduced the Poisson bracket
\be
\{f,g\} = I^{AB}\,\p_Af\wedge\p_Bg
\ee
in terms of the dual $I^{AB}=\half \varepsilon^{ABCD}I_{CD}$ of the infinity twistor. This action will give rise to a theory of self-dual higher-spin gravity on spacetime via the CR-structure deformation 
\begin{equation}
    \dbar_b^\h f:= \dbar_b f + \{\h,f\}\, . 
\end{equation}
On-shell, as discussed in \cite{Mason:2007ct}, this CR structure will be integrable $(\dbar^\h_b)^2=0$. 

The perturbative analysis proceeds in much the same way as that for the Yang-Mills theories above, with Poisson brackets replacing commutators.  The quadratic action hence works as before, and the essential difference is in the vertices where the weight $-2$ of the Poisson structure means that on decomposition into modes $\h_n$ of homogeneity degree $n$ the three-point vertices become 
\begin{equation}
    \frac23 \sum_{l+m+n=-2} \h_l \wedge \{\h_m,\h_n\}\, .
\end{equation}
It follows that in a three-point vertex we must now have that the helicities $h_i$ sum to 2 instead of 1 in the Yang-Mills case; again these will all result in $\overline{\text{MHV}}$ amplitudes.  In flat space, we obtain  \eqref{3pt-mom} but now with $a_1=2h_2+2h_3-2$ and cyclic.  Spacetime formulae analogous to  \eqref{spt--+} and \eqref{spt++-} can be obtained using the local twistor techniques of \cite{Mason:1987,Mason:1987a}.  They still require not only the $(n-1)^\text{th}$ potential $\rho_{\dot\alpha}^{\alpha_2\ldots \alpha_n}$ as above but also the next potential down 
\begin{equation}
    h_{\dot\alpha \dot\beta}
    ^{\alpha_3\ldots \alpha_n}= \nabla_{\alpha_1(\dot\alpha}\rho_{\dot\beta)}^{\alpha_2\ldots \alpha_n}\, .
\end{equation}
which for $n=2$ gives the linearized metric.  It follows from \cite{Mason:1987a} with Einstein infinity twistor that the ASD spin connection $\gamma_{\dot\alpha\alpha_1\alpha_2\alpha_3}=\Lambda \rho_{\dot\alpha\alpha_1\alpha_2\alpha_3}$. 
%as follows from the non-degenerate infinity-twistor analogue of the formulae in \cite{Mason:1987a}.
This leads to  the spin $(-2,2,2)$ vertex
\begin{equation}
       \frac{1}{\Lambda} \int \d^4\mu_\M  \,\psi_{\alpha_1\ldots \alpha_4} \left(\Lambda h^{\alpha_1\alpha_2}_{\dot\alpha\dot\beta}h^{\alpha_3\alpha_4\dot\alpha\dot\beta}+\gamma_{\dot\alpha\beta}^{\alpha_1\alpha_2}\gamma^{\alpha_3\alpha_4\beta\dot\alpha} \right) \,. \label{SDE++-}
\end{equation}
This vertex can be understood as part of a BF Einstein action of MacDowell-Mansouri type, cf.\ \cite{Durka:2011yv,Freidel:2012np}. 
It easily extends to helicities $(-2-l-m,2+l,2+m)$ for $l,m\geq -2$. 

\section{Chiral Moyal higher-spin gravity}\label{Moyal-defs}

The above theories are both contractions of Moyal deformed theories  generalising \cite{Strachan:1992em,Bu:2022iak,Monteiro:2011pc} along the lines of  \cite{Tran:2022tft}. We now discuss the Moyal deformations labeled chs$(\al)$ and gl-chs$(\al)$ in table 1 of \cite{Monteiro:2022xwq}. Start from  our Poisson bivector \begin{equation}
    \Pi:=I^{AB}\p_A \wedge\p_B\, ,
\end{equation} 
the $*$-product arises from its exponentiation 
\begin{equation}
    a*b:=a\exp (\alpha \stackrel{\leftrightarrow}\Pi) b
\end{equation}  
 with deformation parameter $\alpha$. We then deform the cubic Chern-Simons interaction term to 
 \begin{equation}
     \tr\, a^3\rightarrow\tr \left(a\wedge(a*a)\right)\, ,
\end{equation}
where the wedge product between forms is assumed in addition to the $*$-product. 
%Multiple integration by parts yields $\tr (a \wedge(a*a))$ so 
The resulting equations of motion are as in \eqref{YM-eom} but with multiplication replaced  by  the noncommutative $*$-product.  They reduce to  self-dual higher-spin Yang-Mills \eqref{YM-eom} at $\alpha=0$.  Self-dual higher-spin gravity then arises from the GL$(1)$ case by rescaling to pick out the interaction at $O(\alpha)$.

The corresponding 3-point vertices can now be expanded in $\alpha$ and decomposed into modes. The homgeneity degree $-2$ of $\Pi$ shifts the weights by $2r$ in the $\alpha^r$ part of the cubic term giving  
\begin{equation}
 \int \D^3Z \sum_{l+m+n=2r-4}   \tr \,\frac{\alpha^r}{r!}\left(a_l\wedge(a_m\stackrel{\leftrightarrow}\Pi^ra_n)\right)\, .
\end{equation}
  This gives  3-point  vertices with coefficient $\alpha^r/r!$  for  helicities 
\begin{equation}
    h_1+h_2+h_3=r+1\, ,
\end{equation}
now allowing all $+$ three-point vertices for larger $r$.
It is somewhat onerous to find the spacetime formulae for these higher 3-point vertices, although \eqref{SDE++-} provides an $r=1$ example. The momentum space version is easy in flat space because then  the Poisson structure is simply 
\begin{equation}
    \Pi= \varepsilon^{\dot\alpha\dot\beta}\frac{\p}{\p\mu^{\dot\alpha}}\wedge\frac{\p}{\p\mu^{\dot\beta}}\, .
\end{equation} 
Thus its action on momentum eigenstates \eqref{mom-estate} simply introduces additional square-bracket factors into the momentum space 3-point formula. We see that  \eqref{3pt-mom} remains valid but now with $a_1=2h_2+2h_3-r-1$ and cyclic.

%Together with the previous quadratic part, this can be recognized as part of a chiral Plebanski-like Lagrange multiplier action for SD Einstein gravity
%\begin{equation}
%    S[\psi_{\alpha\beta},\Sigma^{\alpha\beta}, \gamma_\alpha^\beta]=\frac{1}{\Lambda}\int \psi_{\alpha\beta}\wedge (R^{\alpha\beta} -\Lambda\Sigma^{\alpha\beta})\label{SDE-pleb}
%\end{equation}
%where $\Sigma^{\alpha\beta}=e^{\alpha}_{\dot\alpha}\wedge e^{\beta\dot\alpha}$ are the ASD 2-forms of a metric, $\psi_{\alpha\beta}=\psi_{(\alpha\beta)} $ a triple of 2-forms, and $R_{\alpha\beta}$ the curvature of an $SL(2,\C)$ connection.\footnote{This follows as the equations of motion give $R^{\alpha\beta}=\Lambda\Sigma^{\alpha\beta}$ so that $D\Sigma^{\alpha\beta}=0$ from the Bianchi identity, giving that the connection is the ASD spin connection of a SD Einstein metric. The variation of the metric gives that $\psi_{\alpha\beta}=\psi_{\alpha\beta\gamma\delta}\Sigma^{\gamma\delta}$ with $\psi_{\alpha\beta\gamma\delta}$ totally symmetric. The variation of the connection then gives that $D\psi_{\alpha\beta}=0$ for the linear ASD field on the SD Einstein background.}

%%%%%%%%%%%%%%%%%%%%%%%%%%%%%%%%%
%%%%%%%%%%%%%%%%%%%%%%%%%%%%%%%%%

\section{Further developments}

The geometrization of spin implicit in the phase of a twistor has always been suggestive of unified theories with arbitrary higher-spins. We have given three such models,  with the first two being scaling limits of the most general Moyal deformed model. %(following the naming convention suggested by  Evgeny Skvortsov).

There are many possible variants of the models discussed here.   
We can introduce supersymmetry by working on supertwistor spaces $\CP^{3|\mathcal{N}}$ with $\cN$ fermionic variables carrying the same phase weights, as in \cite{Boels:2006ir} where they naturally decompose into BF theories. Higher-spin conformal (super-)gravity will be similarly accessible following \cite{Berkovits:2004jj,Adamo:2016ple}. Often one only wishes to keep the integer spin modes: this can be done by imposing $\Z_2$ invariance $a(-Z)=a(Z)$, etc. We can also add fermionic terms $\psi(\dbar_b+a)\psi$ to the 7D action and impose $\psi(-Z)=-\psi(Z)$ to obtain half-integer spin fermions on spacetime without necessarily requiring supersymmetry.  
Models arising from twisted supergravity  \cite{Costello:2016mgj,Raghavendran:2021qbh} can also be considered directly in one of our 7-dimensional twistor spaces, giving an approach to the quantisation of these theories.

Non-chiral higher-spin theories should follow from  similar constructions on ambitwistor space  generalising the Chern-Simons actions of \cite{Mason:2005kn}.  These already encode full (possibly supersymmetric) Yang-Mills theories in terms of CR holomorphic Chern-Simons theory on CR submanifolds of \emph{ambitwistor space}, the quadric inside the cartesian product of twistor space with its dual $\PT\times \PT^*$.  These naturally lift to CR-holomorphic Chern-Simons theories for higher-spin theories defined on the $S^1$ bundle over $\PA$ inside  $\cO(1,1)\rightarrow \PT\times \PT^*$ of unit length. 

An alternative approach to finding non-chiral theories is to find natural interaction terms that extend these chiral theories to full higher-spin theories as, for example, introduced in the restricted spin examples of \cite{Mason:2005zm,Boels:2006ir,Sharma:2021pkl}.
%lifting analogues of the interaction term in the Chalmers-Siegel Yang-Mills action \cite{Chalmers:1996rq} to twistor space.  
%In this direction we remark that, as noted,  \eqref{SDE-pleb} admits such a correction on spacetime. in the addition of the term $\frac{1}{2\Lambda} \int \psi_{\alpha\beta}\wedge \psi^{\alpha\beta}$ that gives an extension to full Einstein gravity in a MacDowell-Mansouri form, for related examples see  \cite{Freidel:2012np}.  
%Note however that there are challenges to this programme for gravity even without the higher-spins \cite{Herfray:2016qvg} and the action in \cite{Sharma:2021pkl} remains speculative. 

In a different direction, our theories start from theories that make good sense before decomposition into higher-spin modes.  One can therefore hope to make full nonlinear sense of these theories without decomposing into an infinite series.  One can further hope to extend the standard integrability constructions  of Penrose and Ward \cite{Penrose:1976js,Ward:1977ta,Ward:1980am}. It will be interesting to study extensions also of Lax pairs, Plebanski potentials and the full twistor constructions without a decomposition into modes;  for example, in our harmonic gauge, the negative modes in the Yang-Mills case naturally extend to the complement of the 4-ball $\langle\sigma \hat\sigma\rangle\leq 1$ inside the compactification of the spin space to $\CP^2$, and the positive modes extend over the interior.  A version of the Ward correspondence applied to the exterior yields a holomorphic connection on the interior $\langle\sigma \hat\sigma\rangle\leq 1$ where the positive helicity modes should converge providing a unified structure. 

Lastly, there exist formal arguments showing that our twistor actions for the higher-spin theories in this paper are free of one-loop gauge and gravitational anomalies \cite{Bittleston:2022nfr}. This is made manifest from the fact that they arise as reductions of theories in 7 dimensions, where such anomalies are expected to be absent on general grounds. This should aid us in constructing their celestial chiral algebras along the lines of \cite{Costello:2022wso,Costello:2022upu,Bittleston:2022jeq}.

%%%%%%%%%%%%%%%%%%%%%%%%%%%%%%%%%
%%%%%%%%%%%%%%%%%%%%%%%%%%%%%%%%%

\medskip

\textit{Acknowledgments:} We would like to thank the participants and organizers of the workshop \href{https://web.umons.ac.be/pucg/en/event/workshop-twistors-and-higher-spins/}{``Twistors and higher-spins''} in Mons 2024, for many contributions and comments on this work; in particular we thank Kirill Krasnov and Evgeny Skvortsov for generously explaining many details of their work and providing helpful comments, alongside Tim Adamo, Roland Bittleston, Claude Lebrun, Arthur Lipstein,  Ricardo Monteiro, David Skinner, and Tung Tran.  We further thank Tim Adamo, Ricardo Monteiro and Evgeny Skvortsov for detailed comments that have improved the draft.  L.M.\ is supported by a grant from the Simons Foundation, `the Simons Collaboration on Celestial Holography' MP-SCMPS-00001550-08 and the STFC consolidated grant ST/X000494/1. A.S.\ is supported by the Gordon and Betty Moore Foundation and the John Templeton Foundation via the Black Hole Initiative.

\bibliographystyle{JHEP}
\bibliography{refs}

\end{document}